\documentclass[12pt]{article}
\usepackage{amssymb}
\usepackage[dvips]{epsfig}
\usepackage{cite}

\renewcommand{\theequation}{\thesection.\arabic{equation}}
\newcommand {\Was}{W\c as}

\newcommand {\KKMC}{\hbox{${\cal KK}$}\ MC}

\newcommand {\eqn}[1]{(\ref{#1})}
\newcommand {\Sp}{\hbox{Sp}}

\newcommand {\fhat}{\hbox{$\hat{f}$}}
\newcommand {\ftilde}{\hbox{$\tilde{f}$}}
\newcommand {\ubar}{\hbox{$\overline{u}$}}
\newcommand {\vbar}{\hbox{$\overline{v}$}}
\newcommand {\pslash}{\hbox{$\not\hbox{\kern-2.3pt $p$}$}}
\newcommand {\kslash}{\hbox{$\not{k}$}}
\newcommand {\epslash}{\hbox{$\not{\epsilon}$}}
\newcommand {\bigM}{\hbox{${\rm\bf M}_1$}}

\newcommand {\MISR}[2]{{\cal M}^{\rm ISR (#2)}_{#1}}
\newcommand {\MFSR}[2]{{\cal M}^{\rm FSR (#2)}_{#1}}
\newcommand {\dsISR}[2]{{d\sigma}^{\rm ISR (#2)}_{#1}}
\newcommand {\fbar}{\overline{f}}
\newcommand {\Ibar}{\overline{I}}
\newcommand {\Rbar}{\overline{R}}
\newcommand {\hvec}[6]{h\left[{#1\atop #4} {#2\atop #5} {#3\atop #6}\right]}
\newcommand {\bigMbar}{\hbox{$\overline{\rm\bf M}_1$}}
\newcommand {\BYFS}{B_{\rm YFS}}
\newcommand {\umf}{\mathfrak{u}}

\newcommand {\avg}[1]{\langle #1 \rangle}
\title{Exact Differential ${\cal O}(\alpha^2)$ Results for Hard Bremsstrahlung
	in $e^+e^-$ Annihilation to Two Fermions At and Beyond LEP2 Energies
\thanks{Work supported in part by the US DoE, contract 
	DE-FG05-91ER40627, by the Polish Government, grants KBN 2P03B08414
	and KBN 2P03B14715, and by the US-Poland Maria Sklodowska-Curie 
	Fund II PAA/DOE-97-316.}
}
\author{
S.\ Jadach\thanks{Permanent address: Institute of Nuclear Physics,
	ul. Kawiory 26a, PL 30-059 Cracow, Poland},\\
	{\normalsize\em CERN, Geneva 23, Switzerland},\\
M. Melles,\\ {\normalsize\em Paul-Scherrer-Institut,
	Wurenlingen und Villigen, Switzerland},\\
B.F.L.\ Ward\thanks{Permanent Address: Department of Physics and Astronomy, 
	University of Tennessee, Knoxville, TN 37996-1200, USA},\\
        {\normalsize\em Werner-Heisenberg-Institut,}\
	{\normalsize\em Max-Planck-Institut fur Phyzik, Munich, Germany}\\
S.\ A.\ Yost\\ {\normalsize\em Department of Physics and Astronomy, 
	University of Tennessee,}\\
	{\normalsize\em Knoxville, Tennessee 37996-1200, USA}
} 
\date{\normalsize July 2001\\[10pt] \sf MPI-PhT-2001-22,
	\quad UTHEP-01-0701}
\begin{document}
\maketitle
\thispagestyle{empty}
\begin{abstract}
\baselineskip=0.55cm
We present the exact ${\cal O}(\alpha)$ correction to the process $e^+ e^-
\rightarrow f\bar{f} + \gamma$, $f\neq e$, for ISR$\oplus$FSR 
at and beyond LEP2 energies.  We give explicit formulas for the completely
differential cross section.  As an important application, we compute
the size of the respective sub-leading corrections of ${\cal O}(\alpha L)$
to the $f\bar{f}$ cross section, where $L$ is the respective
big logarithm in the renormalization group sense so that it is
identifiable as $L = \ln |s|/m_e\!^2$ when $s$ is the squared 
$e^+e^-$ cms energy.  Comparisons are made with the available literature.
We show explicitly that our results
have the correct infrared limit, as a cross-check.  Some comments are
made about the implementation of our results in the framework of the
Monte Carlo event generator \KKMC.
\end{abstract}

\addtolength{\topmargin}{1.5cm}
\addtolength{\textheight}{-1cm}
\newpage
\baselineskip=0.6cm

Currently, the final LEP2 data analysis is in its beginning stages,
and the desired total precision tags on the important
LEP2 physics processes $e^+ e^- \rightarrow f\bar{f}$,
$f\neq e$, are already called out in the LEP2 MC Workshop in
Ref.~\cite{lep2wkshp:2000}.  It has been demonstrated 
in Ref.~\cite{lep2wkshp:2000} that the Monte Carlo (MC) 
event generator program
${\cal KK}$~\cite{kkmc:2001}, hereafter referred to as \KKMC\,, 
and the semi-analytical program 
ZFITTER~\cite{bardin} realize these precisions ($.2 - 1\%$) in most channels 
for inclusive cross sections and that for the fully differential
distributions, the \KKMC\ again meets most of the requirements for the LEP2
final data analysis.  In this paper, we present exact
results on the ${\cal O}(\alpha)$ correction to the single
hard bremsstrahlung processes $e^+ e^- \rightarrow f\bar{f}+\gamma$,
$f\neq e$. This correction is an important contribution to the differential 
distributions as they are realized in the \KKMC\ which allows the
very demanding precisions just cited to be achieved.

Specifically, the exact results for the ${\cal O}(\alpha)$ corrections to
$s$-channel annihilation hard bremsstrahlung processes under study here were 
also considered in Refs.~\cite{berends,in:1987}. We differ from these results
as follows. Concerning Ref.~\cite{berends}, the entire result
was given only for the case in which the photon angle variables
are all integrated out; here, we give the fully differential results.
With regard to Ref.~\cite{in:1987}, the completely differential
results were given as well but the mass corrections were omitted.
In our work, the masses of the electrons and positrons and the 
masses of the final state fermion and anti-fermion are
exactly taken into account in contrast to the literature. 
Thus, by comparing with the two calculations in 
Refs.~\cite{berends,in:1987} as we do here, we
get a measure of the size of the mass corrections as well as
cross checks on both our differential and our integrated results.

Our work is organized as follows. In Section \ref{preliminaries}, we set our 
notational conventions. In Section \ref{exact-virtual},
we present our exact amplitudes for the ${\cal O}(\alpha)$ virtual corrections
to initial-state and final-state real radiation. 
In Section \ref{cross-section}, we derive the 
differential cross-sections corresponding to these amplitudes in a form 
useful for comparisons.  In Section \ref{comparisons}, we compare these 
results with those in Refs.~\cite{berends,in:1987}
while illustrating our results as they are used in the \KKMC\ in 
Ref.~\cite{kkmc:2001}. Section \ref{conclusions} contains our summary 
remarks. The Appendix contains technical details about the
scalar integrals.

\section{Preliminaries}
\label{preliminaries}
\setcounter{equation}{0}
In this section we set our notational conventions. We will use the 
conventions of Refs.~\cite{kkmc:2001,gps:1998,CEEX} for our spinors.
These conventions are based on the Kleiss-Stirling~\cite{KS} Weyl spinors
augmented as described in Refs.~\cite{kkmc:2001,gps:1998,CEEX}
with the rules for controlling their complex phases, or equivalently,
the three axes of the fermion rest frame in which the spin of that fermion is
quantized. We sometimes refer to this fermion rest frame as the
global positioning of spin (GPS) frame and to the rule for determining
it as the GPS rule. The resulting conventions for the fermion spinors are then 
called the GPS spinor conventions. See Refs.~\cite{kkmc:2001,gps:1998,CEEX}
for more details. 
Let us now turn to the kinematics.

\begin{figure}[!ht]
\begin{center}
   \IfFileExists{graphs.eps}{
	\epsfig{file=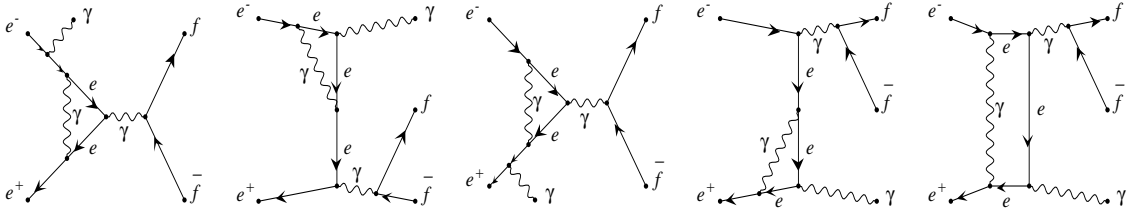,width=15cm}
	}{
	\message{Skipping <graphs.eps>}
	\hbox{[ Figure graphs.eps not included ]}
	}
\end{center}
\vspace{-5mm}
\footnotesize\sf
\caption{
Representative initial state radiation graphs for 
$e^+ e^- \rightarrow f\overline{f}$ 
with one virtual and one real photon, for $f\neq e$.
\label{fig:graphs}
}
\end{figure}

The process under discussion here is shown in Fig.\ \ref{fig:graphs}, the 
one-loop virtual correction to the hard bremsstrahlung process
$e^- e^+ \rightarrow f\bar f +\gamma$, for $f\neq e$. We will treat
both the initial state radiation (ISR) case and the final state
radiation (FSR) case.
We denote the four momenta and helicity of the $e^-$, $e^+$, $f$, and $\bar f$
as $p_j$ and $\lambda_j$,\ $j=1,...,4$, respectively. We denote the 
charge of $f$ by $Q_f$ in units of the positron charge $e$. The rest mass of 
fermion $f$ is denoted by $m_f$. The photon momentum and helicity will
be denoted by $k$ and $\sigma$. With our GPS conventions for spinors,
we induce the following polarization vectors for photons:
\begin{equation}
\label{gmapol}
  (\epsilon^\mu_\sigma(\beta))^*
     ={\bar{u}_\sigma(k) \gamma^\mu u_\sigma(\beta)
       \over \sqrt{2}\; \bar{u}_{-\sigma}(k) u_\sigma(\beta)},\quad
  (\epsilon^\mu_\sigma(\zeta))^*
     ={\bar{u}_\sigma(k) \gamma^\mu \umf_\sigma(\zeta)
       \over \sqrt{2}\; \bar{u}_{-\sigma}(k) \umf_\sigma(\zeta)},
\end{equation}
where the auxiliary 4-vector $\beta$ is exploited here to simplify our
expressions as needed. It satisfies $\beta^2=0$. The second choice with 
$\umf_\sigma(\zeta)$, as defined in Ref.~\cite{kkmc:2001},
is already an example of this exploitation -- it
often leads  to simplifications in the resulting photon emission amplitudes.

The calculations which we present have been done using the program
FORM of Ref.~\cite{form}. For the $t$-channel case, we presented
similar results in Ref.~\cite{virtual} in connection with
the respective ${\cal O}(\alpha^2)L$ corrections needed for the
$0.061\%\ (0.054\%)$ total precision tag achieved in 
Ref.~\cite{bhl4} for the LEP1/SLC luminosity process in the 
Monte Carlo event generator BHLUMI4.04 in Ref.~\cite{bhl4.04}. Just as in the 
latter case, here a considerable effort is needed to simplify our initial raw
FORM output in order to make a practical application
of the respective results in the context of a Monte Carlo environment
such as the \KKMC\ in Ref.~\cite{kkmc:2001}. As in Ref.~\cite{virtual},
we only present the final simplified expressions in this paper for the
sake of clarity.

Our metric is that of Bjorken and Drell in Ref.~\cite{bj-drell}
and we effect our gauge invariant calculation in the 't Hooft-Feynman
gauge.  With these preliminary remarks, we turn now in the next section to 
the calculation of our process of interest.

\section{Exact Results on the Virtual Correction to
\hbox{$e^+e^-\rightarrow$} \hbox{$f\bar f+\gamma$,} 
\hbox{$f\neq e$}} \label{exact-virtual}
\setcounter{equation}{0}

In this section we calculate the exact virtual correction to 
$e^+e^-\rightarrow f\bar f+\gamma$, $f\neq e$. We proceed in analogy
with our results on the virtual correction
for the $t$-channel dominated low angle Bhabha
scattering process with a single hard bremsstrahlung in Ref.~\cite{virtual}.

Specifically, we express the exact amplitude for one real and one
virtual photon emitted from the electron lines in the process
\hbox{$e^+ e^- \rightarrow f\bar f + \gamma$} using the
GPS conventions \cite{gps:1998,CEEX,kkmc:2001}. 
In Ref.~\cite{virtual}, the corresponding $t$ channel result was
obtained for electron line emission. Here, from the latter result,
we first obtain the respective initial-state $s$ channel result by crossing 
the outgoing electron line with the incoming positron line, and
replacing the respective final state by $f\bar f$, while adding also 
$Z$ boson exchange.  The results are translated into GPS conventions.
Then, in subsection \ref{form-factors},
we provide the detailed form factors appearing in the initial state
amplitudes.  The corresponding final state amplitudes are presented
in subsections \ref{final-state} and \ref{finalstate-formfactors}.

\subsection{ ISR $s$-Channel Exact Result }
\label{magic}

In this subsection, we define notation and set up the exact contribution
for one real photon and one virtual
photon emitted from the electron lines in the process 
\hbox{$e^+ e^- \rightarrow f \bar f + \gamma$}. 
The amplitude for real plus virtual photon emission from the
initial state may be written
\begin{equation} \label{iniamp}
\MISR{1}{1} = {Q_e^2 e^2\over 16\pi^2} \MISR{1}{0}\ (f_0 + f_1 I_1 + f_2 I_2)\ ,
\end{equation}
where the real photon emission amplitude is $\MISR{1}{0}$, and the factors
$I_{0,1,2}$ contain spinor dependence. They will be specified in the
next section.  

In GPS conventions, the amplitude $\MISR{1}{0}$ for the
initial state radiation of a single photon is given by
\begin{eqnarray}\label{isramp}
\MISR{1}{0}\left[{p\atop\lambda}\right]
&=& {eQ_e\over 2kp_1} \vbar(p_2,\lambda_2) \bigM
(\pslash_1 + m - \kslash) \epslash_\sigma^\star u(p_1,\lambda_1)
\nonumber\\
&+& {eQ_e\over 2kp_2} \vbar(p_2,\lambda_2) \epslash_\sigma^\star
(-\pslash_2 + m + \kslash) \bigM u(p_1,\lambda_1)\ ,
\end{eqnarray}
where
\begin{equation}\label{Mdef}
\bigM = ie^2 Q_f \sum_{B=\gamma,Z\atop\lambda,\mu =
\pm 1} \left( {\ubar(p_3,\lambda_3)\gamma_\sigma
g_\lambda^{f,B}\omega_\lambda v(p_4,\lambda_4)
\over s'- M_B^2 + i \Gamma_B s'/M_B}\right)\
\gamma^\sigma g_\mu^{e,B} \omega_\mu \end{equation} is the
annihilation scattering spinor matrix.

The form factors may be obtained from the corresponding $t$-channel result in
Ref.~\cite{virtual} for electron line emission. 
Specifically, the $s$ channel result 
can be obtained  by crossing the outgoing electron line with the 
incoming positron line, and replacing the final state by $f\bar f$. 
We also include the effects of $Z$ exchange in the $s$ channel.

Our previous calculations of $t$-channel bremsstrahlung\cite{virtual} 
used the Chinese Magic conventions\cite{xuzhangchang} for the photon
polarizations. The GPS version of the magic photon polarization vector is 
related to the Chinese Magic conventions\cite{xuzhangchang} by
\begin{equation} \label{polarization}
\epsilon^{\star \rm GPS}(k, \beta, \sigma) = \sigma \epsilon^{\rm
Chinese}(k, \beta, -\sigma) .
\end{equation}
The purpose of this change is to recover the more standard 
convention of defining photon polarization in terms of incoming states.
The choice of magic polarization vector affects the amplitude
\eqn{iniamp} only through the definition of $\MISR{1}{0}$.
The remaining factors may thus be obtained directly by crossing from our
previous $t$-channel results. 

The magic choice of auxiliary vector for initial state radiation is
$\beta = \hvec{0}{p_2}{p_1}{\sigma}{\lambda_1}{\lambda_2}$,
with the definition
\begin{eqnarray}\label{hdef}
\hvec{q_0}{q_1}{q_2}{\mu_0}{\mu_1}{\mu_2} &=& \left\{
\begin{array}{ll}
q_0\\
q_1\\
q_2
\end{array} \right\}\quad\hbox{if}\quad\left\{
\begin{array}{ll}
\mu_1 = \mu_2\\
\mu_0 = \mu_1 = -\mu_2\\
\mu_0 = \mu_2 = -\mu_1
\end{array} \right. .
\end{eqnarray}

Using the magic polarization vector in \eqn{iniamp} and
neglecting fermion masses gives
\begin{equation}\label{isrm0}
\MISR{1}{0} = iQ_e\sigma e^3 G_{\lambda_1,\lambda_3}(s')  I_0
{2s_{-\sigma}(p_3,p_4)\over s_\sigma(p_1,k) s_\sigma(p_2,k)} ,
\end{equation}
where the photon-$Z$ propagator is 
\begin{equation}\label{prop}
G_{\lambda,\mu}(s') = \sum_{B=\gamma,Z} {g_\lambda^{e,B}
g_\mu^{f,B} \over
s' - M_B^2 + i \Gamma_B s'/M_B} , 
\end{equation}
and 
\begin{equation}
\label{I0}
I_0 = -\sqrt{2} \lambda_1 \lambda_3
s_{\sigma}^2 \left(\hvec{0}{p_2}{p_1}{\sigma}{\lambda_1}{\lambda_2}, 
\hvec{0}{p_3}{p_4}{\sigma}{\lambda_3}{\lambda_4}\right) .
\end{equation}
We now turn to calculating the form factors and spinor factors.

\subsection{Initial State Form Factors}
\label{form-factors}

It remains to describe the form factors and spinor factors needed to compute 
$\MISR{1}{1}$.  The spinor factors $I_{1,2}$ are given by 
\pagebreak[2]
\begin{eqnarray}
\label{I1}
I_1 &=& \sqrt{2}\lambda_1 s_{-\lambda_1}(p_1,k)
s_{\lambda_1}(p_2,k)\nonumber\\\nopagebreak
& &\quad\times
{s_{-\lambda_1}(p_4,p_1)s_{\lambda_1}(p_1,p_3)
- s_{-\lambda_1}(p_4,p_2)s_{\lambda_1}(p_2,p_3)
\over s_{-\sigma}(p_1,p_2) s_{-\sigma}(p_3,p_4)
s_{-\lambda_1}(p_4,p_2)s_{\lambda_1}(p_2,p_3) I_0} ,\\\pagebreak[2]
\label{I2}
I_2 &=& {\sqrt{2}\sigma s_{-\lambda_1}(p_1,k)
s_{\lambda_1}(p_2,k)
s_{-\lambda_3}(p_4,k) s_{\lambda_3}(p_3,k) \over
s_{-\sigma}(p_1,p_2) s_{-\sigma}(p_4,p_3) I_0} 
\end{eqnarray}
where the spinor product is 
$s_{\lambda}(p, q) = {\bar u}_{-\lambda}(p) u_{\lambda}(q)$. The factors
$I_{1,2}$ are crossed versions of 
\hbox{$({\cal I}_1 \pm {\cal I}_2)/2{\cal I}_0$} in Ref.~\cite{virtual}.

We will begin by writing the dominant term $f_0$. Expressions can be found in
Ref.~\cite{in:1987} for all of the scalar integrals needed for the form 
factors, which were previously calculated using the FF package~\cite{form},
which implements the
methods of Ref.~\cite{scalar}. The integrals in Ref.~\cite{in:1987} are not 
quite adequate, because of the possibility that $r_i < m_e^2/s$. However,
it was possible to analytically continue when necessary, and to reproduce 
the numerical results of the FF package. Thus, an expression for the form 
factors in terms of logarithms and dilogarithms is now available. Details 
on the $s$ channel version of the scalar integrals used in Ref.~\cite{virtual} 
may be found in the Appendix.

For $\sigma = \lambda_1$, using $r_i = 2p_i\cdot k/s$,
\begin{eqnarray}
\label{f0}
f_0 &=& {4\pi}\,{\BYFS}(s,m_e) + 2\left(L - 1 - i\pi\right)
+ {r_2\over 1-r_2}\nonumber\\
&+& {(r_1+r_2) \over (1-r_2) r_1} R(r_1,r_2) +
R(r_2,r_1)\nonumber\\
&+& \left\{3 + {2r_2 \over (1-r_2)(r_1+r_2)}\right\}
\ln\left(1 - r_1-r_2\right) \nonumber\\
&-& {r_2(2+r_1)\over(1-r_1)(1-r_2)}
\left\{\ln{(1-r_1-r_2)\over r_2}
- i\pi\right\}
\end{eqnarray}
with $L = \ln(s/m_e^2)$, the infrared YFS factor
\begin{equation}
\label{BYFS}
4\pi\,{\BYFS}(s,m) = \left(4\ln{m_0\over m} + 1\right)
(\ln{s\over m^2} - 1 - i\pi) - \ln^2({s\over m^2}) - 1
+ {4\pi^2\over 3} + i\pi(2\ln{s\over m^2} - 1)
\end{equation}
and \begin{eqnarray}
R(x,y) &=& \ln^2(1-x) + 2\ln(1-x)
\left\{\ln\left({y\over 1-x}\right) + i\pi\right\}\nonumber\\
&+& 2\,\Sp(x+y) - 2\,\Sp\left({y\over 1-x}\right) \label{Ra2}\\
&=& \ln^2(1-x) + 2\ln(1-x)
\left\{\ln\left({y\over1-x-y}\right) + i\pi\right\}\nonumber\\
&+& 2\ln\left({y\over x+y}\right)\ln\left({1-x-y\over 1-x}
\right) - 2\,\Sp\left({x\over x+y}\right)\nonumber\\
&+& 2\,\Sp(x) + 2\,\Sp\left({(1-x-y)x\over (x+y)(1-x)}\right) .
\label{Rb2}
\end{eqnarray}
The second expression is preferred for calculating $R(x,y)/x$ when
$x$ may be small.  For $\sigma = -\lambda$, $r_1$ and $r_2$ are interchanged 
in \eqn{f0}.

The coefficients of the spinor terms in \eqn{iniamp} are, for 
$\sigma = \lambda$, 
\begin{eqnarray}
\label{f1}
f_1 &=& {(r_1-r_2)\over 2(1-r_1)(1-r_2)}
        + {r_2(1-r_1-r_2)\over r_1(1-r_2)(r_1+r_2)}
        \ln(1 - r_1 - r_2)\nonumber\\
    &+& {r_1+r_2\over 2r_1} \left\{{1-r_1-r_2\over r_1(1-r_2)} +
        {1\over2} \delta_{\sigma,1}\right\} R(r_1,r_2) + {r_1+r_2\over 4r_2}
        \delta_{\sigma, -1} R(r_2,r_1) \nonumber\\
    &+& {1-r_1-r_2\over(1-r_1)(1-r_2)}\left\{{r_1+r_2\over 2(1-r_1)} -
         {1\over 1-r_1} - {r_2\over r_1}\right\}\nonumber\\
    & & \qquad\qquad\qquad\times \left(
         \ln{1-r_1-r_2\over r_2} - i\pi\right)
\end{eqnarray}
and
\begin{eqnarray}
\label{f2}
f_2 &=& 2 - {2-r_1-r_2\over 2(1-r_1)(1-r_2)} \nonumber\\
    &+& {1-r_1-r_2\over r_1(r_1+r_2)} \left({2-r_2\over 1-r_2}\right)
        \ln(1 - r_1 - r_2) \nonumber\\
    &+& {2(1-r_1-r_2)\over r_1+r_2}\left\{1 
        + {1\over r_1+r_2}\ln(1-r_1-r_2)\right\} \nonumber\\
    &+& {(1-r_1-r_2)(2+r_1-r_2)\over 2r_1^2(1-r_2)} R(r_1,r_2)\nonumber\\
    &+& {1\over4} \left(1 - {r_2\over r_1}\right) \delta_{\sigma,1} R(r_1,r_2)
        - {1\over 4} \left(1 - {r_1\over r_2}\right) \delta_{\sigma, -1} 
	R(r_2,r_1)\nonumber\\
    &-& {1-r_1-r_2\over (1-r_1)(1-r_2)}\left({2-r_2\over r_1} 
        + {r_2-r_1\over 2(1-r_1)}\right)\nonumber\\
    & & \qquad\qquad\qquad\times \left(
         \ln{1-r_1-r_2\over r_2} - i\pi\right) .
\end{eqnarray}
The coefficients $f_{1,2}$ are $s$-channel
versions of $({\cal F}_1 \pm {\cal F}_2)/2$ in Ref.~\cite{virtual}.
For $\sigma = -\lambda$, $r_1$ and $r_2$ are interchanged in
\eqn{f1} and \eqn{f2}. 

The leading log limit is obtained by finding which terms give rise to
the leading powers of the `big logarithm' $L$ when the above expressions
are integrated over $r_1$ and $r_2$. These
come from collinear terms where $r_1$ or $r_2$ go to zero. In the collinear
limits, when averaged over the azimuthal angle, only the $f_0$
terms remain to order $L^2$ and $L$, {\em i.e.}, to order NLL.
Using the identities
\begin{eqnarray}
\label{R-identities}
R(0,y) &=& 0 , \nonumber\\
{1\over x}R(x,y) &=& 2\left(1 - {1\over y}\right)
\ln(1-y) - 2\ln y - 2\pi i \hbox{ for } x\rightarrow 0,
\nonumber\\
R(x,y) &=& 2\ln(1-x)\left(\ln y + i\pi\right)
-\ln^2(1-x) + 2 \Sp(x) \hbox{ for } y\rightarrow 0,
\end{eqnarray}
the NLL limit of the form factor $f_0$ is found to be
\begin{eqnarray}
\label{NLL-ini}
f_0^{NLL} &=& 2(L - 1 - i\pi) + 2\ln(1-r_1)(\ln r_2 + i\pi)
+ 2\ln(1-r_2)(\ln r_1 + i\pi)\nonumber\\
&-& \ln^2(1-r_1) - \ln^2(1-r_2) + 3\ln(1-r_1) +
3\ln(1-r_2)\nonumber\\
&+& 2\,\Sp(r_1) + 2\,\Sp(r_2) + {r_1\over
1-r_1}\delta_{\sigma,-\lambda_1}
+ {r_2\over 1-r_2}\delta_{\sigma,\lambda_1} \end{eqnarray}
without mass corrections.

Mass corrections we have calculated primarily without any approximations,
however, in the following we shall present them in the approximation
$m_e<<\sqrt{s}$.
In particular, in this approximation,
we checked by explicit calculation that the
result which we obtain for the mass corrections in fact agrees with that
implied by the prescription in Ref.~\cite{berklei}.
This prescription is valid for the spin-averaged differential distribution
in the limit $m_e<<\sqrt{s}$, but since mass terms are located in the
separate
(helicity conservation violating) spin amplitudes, it is not difficult
to ``undo'' the spin summation.
The technique of Ref.~\cite{berklei} was originally
applied to tree level photon emissions.
Following the Appendix B of Ref.~\cite{virtual} we can apply it also to
our case of emission of one virtual and one real photon.

Taking advantage of the freedom, which we have for presenting
mass terms in the $m_e<<\sqrt{s}$ approximation,
the introduction of the mass correction leads to a replacement
of $f_0$ by $f_0 + f_0^{m_e}$, where
\begin{eqnarray}
\label{mass-ini}
f_0^{(m_e)} &=& {2 {m_e}^2\over s} \left({r_1\over r_2} +
{r_2\over r_1} \right)
{(1-r_1)(1-r_2)\over 1+ (1-r_1-r_2)^2} \nonumber\\
&\times& \left\{f_0 - 4\pi\;{\BYFS}(s',m_e) - 2[L +
\ln(1-r_1-r_2)- 1 - i\pi]
\right\}
\end{eqnarray}
with YFS infrared factor
\begin{eqnarray}
\label{BYFS-ini}
4\pi\,{\BYFS}(s', m_e) &=& 4\pi\,{\BYFS}(s, m_e) +
\ln(1-r_1-r_2)\;
\big(4\ln{m_0\over m_e}\nonumber\\
&-& 2L+1+2\pi i \big) - \ln^2 (1-r_1-r_2) \ .
\end{eqnarray}
Mass corrections first appear at order NLL, and to this order,
\begin{eqnarray}
\label{mass-ini-NLL}
f_0^{(m_e)\ NLL} &=& {2 {m_e}^2\over s} \left({r_1\over r_2} +
{r_2\over r_1} \right)
{(1-r_1)(1-r_2)\over 1+ (1-r_1-r_2)^2} \nonumber\\
&\times& \Big\{\ln(1-r_1-r_2)\;\left(2L-1-2\pi i -
4\ln{m_0\over m_e}\right) + \ln^2 (1-r_1-r_2) \nonumber\\
&+& 2\ln(1-r_1)(\ln r_2 + i\pi) + 2\ln(1-r_2)(\ln r_1 + i\pi)
\Big\}. \end{eqnarray}
Only the LL part of $f_0$ contributes to the mass correction, to order NLL.
The result (\ref{mass-ini}) gives the complete effect of the mass corrections
for the ISR neglecting the terms that are suppressed by higher powers of
$m_e^2/s$ as usual. 

\subsection{Final State Radiation}
\label{final-state}

The amplitudes for final state radiation can be obtained by crossing the
incoming electron with the outgoing $\bar f$, and the incoming positron
with the outgoing $f$. Thus, $p_1 \leftrightarrow -p_4$,
$p_2 \leftrightarrow -p_3$, $\lambda_1 \leftrightarrow -\lambda_4$, and
$\lambda_2 \leftrightarrow -\lambda_3$ in the results of the previous sections.

The final state radiation (FSR) amplitude can be written in analogy with
the ISR result \eqn{iniamp}, 
\begin{equation} \label{finamp}
\MFSR{1}{1} = {Q_f^2 e^2\over 16\pi^2} \MFSR{1}{0}\
(\fbar_0
+ \fbar_1 \Ibar_1 + \fbar_2 \Ibar_2) \end{equation}
The form factors ${\fbar}_{0,1,2}$ and the spinor factors
${\Ibar}_{1,2}$ are final-state analogs of those in the previous
section, and will be defined in the next subsection.

The amplitude $\MFSR{1}{0}$ for the final state radiation of a
single photon can be obtained from the initial state amplitude
$\MISR{1}{0}$ by crossing. Crossing leads to spinors with
negative energy. A consistent choice of branches gives
\begin{equation}
u(-p, -\lambda) = i v(p,\lambda), \qquad
v(-p, -\lambda) = i u(p, \lambda).
\end{equation}
Then we obtain
\begin{eqnarray}\label{fsramp}
\MFSR{1}{0}\left[{p\atop\lambda}\right]
&=& { eQ_f \over 2kp_4 } \ubar(p_3,\lambda_3) \bigMbar
(\pslash_4 - m + \kslash) \epslash_\sigma^\star v(p_4,\lambda_4)
\nonumber\\
&-& {eQ_f \over 2kp_3} \ubar(p_3,\lambda_3)
\epslash_\sigma^\star
(\pslash_3 + m + \kslash) \bigMbar v(p_4,\lambda_4),
\end{eqnarray}
where
\begin{equation}\label{Mbardef}
\bigMbar = ie^2 Q_e \sum_{B=\gamma,Z\atop\lambda,\mu =
\pm 1} \left(
{\vbar(p_2,\lambda_2)\gamma_\sigma
g_\lambda^{e,B}\omega_\lambda u(p_1,\lambda_1)
\over s - M_B^2 + i \Gamma_B s/M_B}\right)\
\gamma^\sigma g_\mu^{f,B} \omega_\mu .
\end{equation}

The magic polarization vector for final state radiation is
$\hvec{0}{p_3}{p_4}{\sigma}{\lambda_3}{\lambda_4}$.
Using this in \eqn{fsramp} gives, in the massless limit,
\begin{equation}
\MFSR{1}{0} = iQ_f\sigma e^3
G_{-\lambda_4,-\lambda_2}(s) I_0
{2s_{-\sigma}(p_3,p_4)\over s_\sigma(p_1,k) s_\sigma(p_2,k)} ,
\end{equation}
with propagator \eqn{prop} and $I_0$ given again by \eqn{I0}.

\subsection{Final State Form Factors}
\label{finalstate-formfactors}

The spinor factors $\Ibar_{1,2}$ appearing in $\MFSR{1}{1}$
are given by \begin{eqnarray}
\label{Ibar1}
\Ibar_1 &=& \sqrt{2}\lambda_4 s_{\lambda_4}(p_4,k)
s_{-\lambda_4}(p_3,k)
\nonumber\\
& &\quad\times
{s_{\lambda_4}(p_1,p_4)s_{-\lambda_4}(p_2,p_4)
+ s_{\lambda_4}(p_1,p_3)s_{-\lambda_4}(p_3,p_2)
\over s_{-\sigma}(p_4,p_3) s_{-\sigma}(p_2,p_1)
s_{\lambda_4}(p_1,p_3)s_{-\lambda_4}(p_3,p_2) I_0} ,\\
\label{Ibar2}
\Ibar_2 &=& {\sqrt{2}\sigma s_{\lambda_4}(p_4,k)
s_{-\lambda_4}(p_3,k)
s_{\lambda_2}(p_1,k) s_{-\lambda_2}(p_2,k) \over
s_{-\sigma}(p_4,p_3) s_{-\sigma}(p_1,p_2) I_0} .
\end{eqnarray}

As before, let \hbox{$r_i = 2p_i\cdot k/s$}. We can
obtain the form factor $\fbar_0$ for \hbox{$\sigma = -\lambda_3$}
by substituting 
\begin{equation}
\label{rsub}
r_1 \rightarrow -r_3/(1-r_3-r_4), \qquad r_2 \rightarrow
-r_4/(1-r_3-r_4)
\end{equation}
in \eqn{f0}, and for \hbox{$\sigma = +\lambda_3$} by interchanging $r_3$ and
$r_4$ in \eqn{rsub}. (Since \hbox{$k^2 = 0$}, we have
\hbox{$r_1 + r_2 = r_3 + r_4$}.)
Then, for $\sigma = \lambda_3$,\pagebreak[2]
\begin{eqnarray}
\label{fbar0}
\fbar_0 &=& 4\pi\,{\BYFS}(s', m_f) + 2\left(L -2\ln{m_f\over
m_e} -1 - i\pi\right)
- {r_3\over 1-r_4} \nonumber\\\nopagebreak
&+& {(r_3+r_4)(1-r_3-r_4) \over r_4(1-r_4)} \Rbar(r_3,r_4)
+ \Rbar(r_4,r_3)\nonumber\\
&-& \left\{1 + {2r_3 (1-r_3-r_4)\over
(1-r_4)(r_3+r_4)}\right\}
\ln\left(1 - r_3-r_4\right) \nonumber\\
&+& {r_3\over 1-r_4}\left\{{3r_4\over 1-r_3} - 2\right\}
\ln{r_3} \end{eqnarray}
with \begin{eqnarray}
\label{Rbar}
\Rbar(x,y) &=& R\left({-y\over 1-x-y}, {-x\over
1-x-y}\right)
\nonumber\\
&=& 2\ln x \ln\left({1-x\over 1-x-y}\right)
+ 2\,\Sp(x) - 2\,\Sp(x+y) .
\end{eqnarray}
The expression \eqn{Rbar} is obtained from \eqn{Ra2} using the
dilogarithm identity
\begin{equation}
\Sp\left({-x\over 1-x}\right) = -\Sp(x) - {1\over2}\ln^2(1-x) .
\end{equation}
The imaginary parts in \eqn{fbar0} were obtained by assuming the $i\pi$
terms in \eqn{f0} came from a small positive imaginary part on $s$ or $s'$.
For $\sigma = -\lambda_3$, $f_0 = f_0(r_4, r_3)$ instead: $r_3$ and $r_4$ are
interchanged.

The coefficients of the spinor terms in \eqn{finamp} are, for
$\sigma = \lambda_3$, \begin{eqnarray}
\label{fbar1}
\fbar_1 &=& {(r_3-r_4)(1-r_3-r_4)\over 2(1-r_3)(1-r_4)} +
{r_3(1-r_3-r_4)\over r_4(1-r_4)(r_3+r_4)} \ln(1 - r_3 -
r_4)\nonumber\\
&-& {r_3+r_4\over 2r_4} \left\{{1-r_3-r_4\over r_4(1-r_4)}
-
{1\over2} \delta_{\sigma,-1}\right\} \Rbar(r_3,r_4)
+ {r_3+r_4\over 4r_3} \delta_{\sigma, 1} \Rbar(r_4,r_3)
\nonumber\\
&+& {(1-r_3-r_4)\over(1-r_3)(1-r_4)}\left\{1 + {r_3\over
r_4} +
{r_3-r_4\over 2(1-r_3)}\right\} \ln r_3 ,
\end{eqnarray}
\begin{eqnarray}
\label{fbar2}
\fbar_2 &=& 2 - {(1-r_3-r_4)(2-r_3-r_4)\over
2(1-r_3)(1-r_4)} \nonumber\\
&-& {1-r_3-r_4\over r_4(r_3+r_4)} \left(2 - {r_3\over
1-r_4}\right)
\ln(1 - r_3 - r_4) \nonumber\\
&-& {2\over r_3+r_4}\left\{1 + {1-r_3-r_4\over
r_3+r_4}\ln(1-r_3-r_4)
\right\} \nonumber\\
&+& {(1-r_3-r_4)(2-r_3-3r_4)\over 2r_4^2(1-r_4)}
\Rbar(r_3,r_4)\nonumber\\
&+& {1\over4} \left(1 - {r_3\over r_4}\right)
\delta_{\sigma,-1}
\Rbar(r_3,r_4) - {1\over 4} \left(1 - {r_4\over r_3}\right)
\delta_{\sigma, 1} \Rbar(r_4,r_3)
\nonumber\\
&-& {1-r_3-r_4\over (1-r_3)(1-r_4)}\left\{{2-r_3\over r_4}
+ {r_4-r_3\over
2(1-r_3)}-2\right\}\ln r_3 .
\end{eqnarray}
For $\sigma = -\lambda_3$, $r_3$ and $r_4$ are interchanged in
\eqn{fbar1} and \eqn{fbar2}.

The NLL limit is obtained as in the initial state radiation case, except that
now the collinear limits are when $r_3$ or $r_4$ become small.  Only
the form factor $\fbar_0$ survives to order NLL, and using the identities
\begin{eqnarray}
\label{Rbar-identities}
\Rbar(x,0) &=& 0 ,\nonumber\\
{1\over y}\Rbar(x,y) &=& {2\over 1-x} \ln x + {2\over x} \ln
(1-x)
\hbox{ for } y\rightarrow 0,\nonumber\\
\Rbar(x,y) &=& -2\ln x \ln (1-y) - 2\,\Sp(y)
\hbox{ for } x\rightarrow 0, \end{eqnarray}
we find
\begin{eqnarray}
\label{NLL-fin}
\fbar_0^{NLL} &=& 4\pi\,{\BYFS}(s', m_f) + 2(L-1-i\pi)
\nonumber\\
&-& 2\ln r_3 \ln(1-r_4) - 2\ln r_4 \ln(1-r_3) \nonumber\\
&-& \ln(1-r_3) - \ln(1-r_4) - 2\,\Sp(r_3) -
2\,\Sp(r_4)\nonumber\\
&-& \delta_{\sigma,\lambda_3} r_3 -
\delta_{\sigma,-\lambda_3} r_4
\end{eqnarray}
without mass corrections.

Spin-averaged mass corrections can be obtained from the initial state case
\eqn{mass-ini} by crossing. The result is that $\fbar_0
\rightarrow \fbar_0 + \fbar_0^{m_f}$, where
\begin{eqnarray}
\label{mass-fin}
\fbar_0^{(m_f)} &=& {2 m_f^2\over s} \left({r_3\over r_4} +
{r_4\over r_3} \right)
{(1-r_3)(1-r_4)\over 1+ (1-r_3-r_4)^2} \nonumber\\
&\times& \left\{f_0 - 4\pi\,{\BYFS}_f(s) - 2(L -
2\ln{m_f\over m_e}
- 1 - i\pi)\right\}
\end{eqnarray}
where
\begin{eqnarray}
\label{BYFS-fin2}
4\pi\,{\BYFS}_f(s) &=& 4\pi\,{\BYFS}(s', m_f) -
\ln(1-r_3-r_4)
\big(4\ln{m_0\over m_e} \nonumber\\
&-& 2L+1+2\pi i \big) + \ln^2 (1-r_3-r_4) \ .
\end{eqnarray}
Again, mass corrections first appear at order $NLL$, and to this
order, \begin{eqnarray}
\label{mass-fin-NLL}
\fbar_0^{(m_f)\ NLL} &=& -{2 m_f^2\over s} \left({r_3\over r_4}
+
{r_4\over r_3} \right) {(1-r_3)(1-r_4)\over 1+ (1-r_3-r_4)^2}
\nonumber\\
&\times& \Big\{\ln(1-r_3-r_4)\left(4\ln{m_0\over m_f}
- 2L + 4\ln{m_f\over m_e}+1+2\pi i\right) \nonumber\\
&-& \ln^2(1-r_3-r_4) + 2\ln(1-r_3)\ln r_4 + 2\ln(1-r_4)\ln r_3
\Big\}.
\end{eqnarray}
The result (\ref{mass-fin}) gives the complete effect of FSR mass
corrections neglecting terms suppressed by higher powers of
${m_f^2\over s}$ as usual.

\section{Differential Cross Section}
\label{cross-section}
\setcounter{equation}{0}

This section translates our amplitudes into differential cross sections,
and sets up comparisons with other related results.
The initial state differential cross section for emitting one real
photon
may be written
\begin{equation}
\label{dsISR10}
{\dsISR{1}{0}\over d^2\Omega dr_1 dr_2} = {1\over 2(4\pi)^4
s'} \sum_{\lambda_i, \sigma} \left| \MISR{1}{0}\right|^2\ ,
\end{equation}
where the summed, squared real photon amplitude leads to
\begin{equation}
\left| \MISR{1}{0}\right|^2 = {Q_e^4 e^6\over s^2 r_1 r_2}
\left[ (t_1^2 + t_2^2) (F_0 - F_1) + (u_1^2 + u_2^2)(F_0 + F_1)
\right]\ , 
\end{equation}
where the invariants $u_i$, $t_i$ may be written 
in YFS3-style~\cite{yfs3-norm,kkmc:2001} effective angle notation:
\begin{eqnarray}
\label{t-u-def}
t_i &= (p_i-p_{i+2})^2 &= - {1\over4}\beta_f\beta_i s(1-r_i)
[2 - \cos(\theta_{1i}) - \cos(\theta_{2i})]\nonumber\\
u_i &= (p_i-p_{j+2})^2 &= - {1\over4}\beta_f\beta_i s(1-r_i)
[2 + \cos(\theta_{1i}) + \cos(\theta_{2i})]
\end{eqnarray}
with $(i,j) = (1,2)$ or $(2,1)$, and \begin{equation}
\beta_f = \sqrt{1 - {4m_f^2\over s'}} , \qquad
\beta_i = \sqrt{1 - {4m_e^2s'\over s^2(1-r_i)^2}} .
\end{equation}
We will be setting these mass factors to unity in the following,
and adding mass corrections at the end via \eqn{mass-ini} or
\eqn{mass-fin}.

The coefficients $F_i$ are defined in terms of the standard vector and
axial vector fermion couplings $V$ and $A$, and \hbox{$X = V^2 + A^2$},
\hbox{$Y = 2VA$} by
\begin{eqnarray}
F_0 &=& X_eX_f\chi_2 + 2Q_eQ_fV_eV_f\chi_1 + Q_e^2 Q_f^2,\nonumber\\
F_1 &=& Y_eY_f\chi_2 + 2Q_eQ_fA_eA_f\chi_1,
\end{eqnarray}
where
\begin{eqnarray}
\chi_1 &=& s'(s'-M_Z^2) [(s'-M_Z^2)^2 +
(s'\Gamma_Z/M_Z)^2]^{-1} , \nonumber\\
\chi_2 &=& {s'}^2 [(s'-M_Z^2)^2 +
(s'\Gamma_Z/M_Z)^2]^{-1}. \end{eqnarray}

The initial state differential cross section for real plus virtual
photon emission may be expressed as
\begin{equation}
\label{dsISR11}
{\dsISR{1}{1} \over d^2\Omega dr_1 dr_2} = {1\over (4\pi)^4
s'}\sum_{\lambda_i, \sigma}{\rm Re}
\left[(\MISR{1}{0})^{*} \MISR{1}{1}\right] . \end{equation}

If is convenient to rewrite \eqn{iniamp} as
\begin{equation}
\label{vdef1}
\MISR{1}{1} = {Q_e^2 e^2\over 16\pi^2}\; v\; \MISR{1}{0}
\end{equation}
in terms of a virtual correction factor
\begin{equation}
\label{vdef2}
v = f_0 + f_1 I_1 + f_2 I_2.
\end{equation}
The differential cross section \eqn{dsISR11} can then be written
in terms of a spin-averaged virtual correction factor $\avg{v}$
times the cross section for pure real initial state radiation:
\begin{equation}
\dsISR{1}{1} = \avg{v}\; \dsISR{1}{0} ,
\end{equation}
where
\begin{eqnarray}
\label{vfac}
\avg{v} &=& {\sum_{\sigma,\lambda_1,\lambda_3} 
    v\; |\MISR{1}{0}|^2 \over
    \sum_{\sigma,\lambda_1,\lambda_3} |\MISR{1}{0}|^2} \nonumber\\
  &=& {\sum_{\lambda_1}\left\{v_+^{\lambda_1}[F_0+F_1+\lambda_1(F_2+F_3)]
        + v_{-}^{\lambda_1}[F_0-F_1+\lambda_1(F_2-F_3)]\right\}\over
     (F_0+F_1)(u_1^2 + u_2^2) + (F_0-F_1)(t_1^2 + t_2^2)} 
\end{eqnarray}
with
\begin{eqnarray}
v_+^\lambda &=& u_1^2 v_{\lambda,\lambda,-\lambda}
            + u_2^2 v_{\lambda,\lambda,\lambda} , \nonumber\\
v_-^\lambda &=& t_1^2 v_{\lambda,-\lambda,-\lambda}
            + t_2^2 v_{\lambda,-\lambda,\lambda} ,
\end{eqnarray}
and
\begin{eqnarray}
F_2 &=& Y_eX_f\chi_2 + 2Q_eQ_fA_eV_f\chi_1,\\
F_3 &=& X_eY_f\chi_2 + 2Q_eQ_fV_eA_f\chi_1.\nonumber
\end{eqnarray}

In the NLL approximation, where the $f_1$ and $f_2$ terms in
\eqn{iniamp} may be neglected, we may use the relation
$\avg{v} = \avg{f_0}$. Then \eqn{vfac} simplifies to 
\begin{eqnarray}
\label{spinavg1}
&\avg{f_0} =
2{(f_0(r_1,r_2) u_2^2 + f_0(r_2,r_1) u_1^2)(F_0+F_1)
+(f_0(r_1,r_2) t_2^2 + f_0(r_2,r_1) t_1^2)(F_0-F_1)
\over
(u_1^2+u_2^2)(F_0+F_1) +
(t_1^2+t_2^2)(F_0-F_1)}\nonumber\\[10pt]
&= 
{F_0[\fhat_0(r_1,r_2)(1+\cos^2\theta_{11}) +\fhat_0(r_2,r_1)(1+\cos^2\theta_{22})]
 + 2F_1[\fhat_0(r_1,r_2)\cos\theta_{11} +\fhat_0(r_2,r_1)\cos\theta_{22}]
\over
F_0[(1-r_1)^2(1+\cos^2\theta_{11}) +(1-r_2)^2(1+\cos^2\theta_{22})] 
+ 2F_1[(1-r_1)^2\cos\theta_{11} +(1-r_2)^2\cos\theta_{22}]}
\end{eqnarray}
with
\begin{equation}
\fhat_0(r_i, r_j) = 2 (1- r_j)^2 f_0(r_i, r_j) .
\end{equation}

If we further drop the dependence on $\theta_{ij}$ in \eqn{t-u-def}, letting
$\theta_{ij}\rightarrow \theta$ for fixed $\theta$, then we get a
simpler approximation. The angle dependence can be factored out
of the cross-section \eqn{dsISR11}, leading to \begin{equation}
\label{dsISR-approx}
{\dsISR{1}{1}\over dr_1 dr_2} = {1\over2}\left({Q_e e^2 \over
2\pi^2}\right)^2
\sigma_0\langle f_0 \rangle H_0(r_1, r_2) .
\end{equation}
with the definition
\begin{equation}
\label{Hdef}
H_0(r_1, r_2) = {1\over 2r_1 r_2} \left[(1-r_1)^2 +
(1-r_2)^2\right] ,
\end{equation}
the total Born cross section \begin{equation}
\label{born}
\sigma_0 = {Q_e^2 Q_f^2 e^4\over 2(4\pi)^2 s'}\int
d^2\Omega
\left({1\over2}F_0(1 + \cos^2\theta) + F_1\cos \theta\right) ,
\end{equation}
and approximate spin-averaged form factor
\begin{equation}
\label{spinavg2}
\langle f_0 \rangle = {(1-r_2)^2 f_0(r_1,r_2) + (1-r_1)^2
f_0(r_2,r_1) \over
(1 - r_1)^2 + (1 - r_2)^2}.
\end{equation}
It can be shown that this approximation is valid to order NLL.

The NLL expression \eqn{NLL-ini} then leads to the spin-averaged form factor
\begin{eqnarray}
\label{NLL-ini-avg}
\langle f_0^{NLL}\rangle &=& 2(L - 1 - i\pi) + 2\ln(1-r_1)(\ln
r_2 + i\pi)
+ 2\ln(1-r_2)(\ln r_1 + i\pi)\nonumber\\
&-& \ln^2(1-r_1) - \ln^2(1-r_2) + 3\ln(1-r_1) +
3\ln(1-r_2)\nonumber\\
&+& 2 \Sp(r_1) + 2 \Sp(r_2) + {r_1(1-r_1)\over 1 + (1-r_1)^2}
+ {r_2(1-r_2)\over 1 + (1-r_2)^2}
\end{eqnarray}
with mass corrections given by \eqn{mass-ini}.
This form is useful for comparison with other results on the
differential cross section, as we will see in the next section.

An analogous expression can be found for the final state emission
cross section. The spin-averaged version of the final state form
factor \eqn{NLL-fin} is
\begin{eqnarray}
\label{NLL-fin-avg}
\fbar_0^{NLL} &=& 4\pi\,{\BYFS}(s', m_f) + 2(L-1-i\pi) -
2\ln r_3 \ln(1-r_4)
\nonumber\\
&-& 2\ln r_4 \ln(1-r_3) - \ln(1-r_3) - \ln(1-r_4) - 2 \Sp(r_3)
- 2 \Sp(r_4)\nonumber\\
&-& {r_3\over 1 + (1-r_3)^2} - {r_4\over 1 + (1-r_4)^2}
\end{eqnarray}
with mass corrections given by \eqn{mass-fin}.

\section{Partly Differential Cross Section and Comparisons}
\label{comparisons}
\setcounter{equation}{0}

We may compare our spin-averaged initial state radiation form factor with
one published in Ref.~\cite{in:1987}. In our notation, this result
may be written as
\begin{equation}
f_{IN} = {\ftilde_{IN}(r_1,r_2) + \ftilde_{IN}(r_2,r_1) \over
(1 - r_1)^2 + (1 - r_2)^2}
\end{equation}
with
\begin{eqnarray}
\ftilde_{IN}(r_1, r_2) &=& (1-r_1-r_2+r_2^2)\left[4\pi{\rm
Re}\,\BYFS(s, m_e) + 2(L - 1)\right]
\nonumber\\
&+& [1+(1-r_2)^2]\left\{ \ln^2(r_1) + \ln^2{1-r_1-r_2\over
1-r_2}
- \ln^2{r_1\over 1-r_1-r_2}\right. \nonumber \\
& & \qquad \qquad \qquad + \left. 2\,\Sp\left({1-r_1-r_2\over
1-r_2}\right)
+ 2\,\Sp(r_1+r_2) - {\pi^2\over 3}\right\}\nonumber\\
&+& {r_1 r_2 (1-r_1-r_2)\over 1-r_2} \left({1\over 1-r_2} + 1
- {3r_2\over r_1}\right) \ln{r_1\over 1-r_1-r_2} \nonumber\\
&-& {2r_1 r_2 (1-r_1-r_2)\over r_1+r_2}
\left({(1-r_1-r_2)\over r_1+r_2}
- {3\over r_1}\right) \ln(1-r_1-r_2) \nonumber\\
&+& r_2(3r_2 + 2r_1)\ln(r_1) - {2r_1 r_2 \over r_1+r_2}
+ {r_1 r_2 \over 1-r_2} + r_2(1-r_2) .
\end{eqnarray}
The NLL limit of this expression may be obtained by summing
the two collinear limits where $r_i \rightarrow 0$ separately.
Carrying this out leads precisely to our spin-averaged NLL expression
\eqn{NLL-ini-avg}. Thus, we agree with Ref.~\cite{in:1987} to
NLL order.  Note that the expression $f_{IN}$ does not include mass
corrections, which were not calculated in Ref.~\cite{in:1987}.

By integrating out the separate dependence on $r_i$ in favor of
the variable\linebreak[1] \hbox{$z = s'/s$} \hbox{$= 1 - r_1 - r_2$},
\linebreak[1] we may obtain a result which can be compared to 
Ref.~\cite{berends}.  To begin, we consider the pure
real photon ISR cross section, and work in the approximation where the
effective angles $\theta_{ij}$ in \eqn{t-u-def} are replaced by a common
angle $\theta$. Then \begin{equation}
\label{realDCS}
{\dsISR{1}{0}\over dr_1 dr_2} = {Q_e^2 e^2\over
4\pi^2}\sigma_0 H_0(r_1, r_2)
\end{equation}
in terms of $H_0$ defined by \eqn{Hdef}. Integrating $r_1$ and
$r_2$ with
the constraint $z = 1 - r_1 - r_2$ gives
\begin{equation}
{\dsISR{1}{0}\over dz} = {Q_e^2 e^2\over 4\pi^2}\sigma_0
\int_{r_0}^{1-z-r_0} dr_1 dr_2 \delta(1-z-r_1-r_2) H_0(r_1,r_2),
\end{equation}
where \hbox{$r_0 = m_e^2(1-z)/s$} is the kinematic minimum
value of $r_1$ or
$r_2$. The result of the integral is, exactly, \begin{equation}
{\dsISR{1}{0}\over dz} = {Q_e^2 e^2\over 4\pi^2}\sigma_0
\left(1+z^2\over 1-z\right)\left[ L - {(1-z)^2\over 1+z^2}\right]
.
\end{equation}

Mass corrections, obtained by the prescription of
Ref.~\cite{berklei},  have the effect of replacing $H_0$ by
\hbox{$H_0 + H_m$} in \eqn{realDCS}, where
\begin{equation}
\label{realDCS-mass}
H_m(r_1, r_2) = - {2m_e^2\over s} \left(
{r_1\over r_2} + {r_2\over r_1} \right) {z\over 1+z^2} H_0(r_1,
r_2) .
\end{equation}
Integrating the mass term gives \begin{equation}
\int_{r_0}^{1-z-r_0} dr_1 dr_2 H_m(r_1, r_2) =
-{2z\over 1-z} + {\cal O}({m_e^2\over s}).
\end{equation}
The total mass-corrected real photon emission cross section is then
\begin{equation}
\label{realDCS-total}
{\dsISR{1}{0}\over dz} = {Q_e^2 e^2\over 4\pi^2}\sigma_0
\left(1+z^2\over 1-z\right)( L - 1) = \delta_1^{H_1}(z)\sigma_0.
\end{equation}
in the notation of Ref.~\cite{berends}, where this result matches
the real part of \eqn{f0}.

The real plus virtual ISR cross section is \begin{equation}
\label{virtualDCSeq}
{\dsISR{1}{1}\over dz} = {1\over 2}\left({Q_e e^2\over
4\pi^2}\right)^2
\sigma_0 \int_{r_0}^{1-z-r_0} dr_1 dr_2 \delta(1-z-r_1-r_2)
\langle f_0 \rangle H_0(r_1, r_2),
\end{equation}
where we use the NLL expression \eqn{NLL-ini-avg} for the virtual form factor,
and will add mass corrections later. Doing the integral and
keeping all infrared terms and terms of order $L^2$ and $L$ gives
\begin{eqnarray}
\label{virtDCS-massless}
{\dsISR{1}{1}\over dz} &=&
{Q_e^2 e^2\over
4\pi^2}\sigma_0\delta_1^{V_1}(s)\left({1+z^2\over 1-z}\right)
\left[L - {(1-z)^2\over 1+z^2}\right] \nonumber\\
&+& {1\over2} \left({Q_e^2 e^2\over 4\pi^2}\right)^2\sigma_0
L
\left(1+z^2\over 1-z\right) \Big\{ - L\ln z + 2\ln z \ln(1-z)
\nonumber\\
&+& 2\,\Sp(1-z) + 3\ln z - \ln^2 z
+ {z(1-z)\over 1 + z^2}\Big\},
\end{eqnarray}
where we use the notation of Ref.~\cite{berends} for
\begin{equation}
\delta_1^{V_1}(s) = {Q_e^2 e^2\over 4\pi^2} \left\{2\pi\BYFS(s,
m_e) + L -1\right\}.
\end{equation}

The mass correction is obtained by multiplying the pure
real mass correction by the single virtual photon form factor evaluated
in terms of $s'$ rather than $s$. Thus, we add to the differential
cross section a mass term 
\begin{equation}
{d\sigma_m\over dr_1 dr_2} = {Q_e^2 e^2\over 4\pi^2}
\delta_1^{V_1}(s') H_m(r_1, r_2)\ \sigma_0,
\end{equation}
where 
\begin{eqnarray}
\delta_1^{V_1}(s') &=& {Q_e^2 e^2\over 4\pi^2}
\left\{2\pi\BYFS(s', m_e)
+ L + \ln z - 1\right\}
\nonumber\\
&=& \delta_1^{V_1}(s) + \left({Q_e e^2\over 4\pi^2}\right) \ln
z
\left( 2\ln{m_0\over m_e} - L + {3\over 2} - \ln^2 z \right) .
\end{eqnarray}
Integrating over $r_1$ and $r_2$ with $z = 1-r_1-r_2$ and keeping only
infrared terms and terms of order $L$, we obtain
\begin{equation}
{d\sigma_m\over dz} = - {Q_e^2 e^2\over 4\pi^2} {2z\over
1-z}\sigma_0
\Big\{\delta_1^{V_1}(s) - {Q_e^2 e^2\over 4\pi^2}L\ln z\Big\}
.
\end{equation}
Adding the mass corrections and using the notation of \eqn{realDCS-total}
gives the complete real plus virtual cross section at order NLL,
\begin{eqnarray}
&{\dsISR{1}{1}\over dz} = \sigma_0 \delta_1^{V_1}(s)
\delta_1^{H_1}
+ {1\over 2}\left({Q_e^2 e^2\over 4\pi^2}\right)^2\sigma_0 L
\Big\{
\left(4z\over 1-z\right) \ln z + z& \nonumber\\
&+ \left(1+z^2\over 1-z\right) \big[ -L\ln z + 2\ln z \ln(1-z) +
3\ln z
- \ln^2 z + 2\,\Sp(1-z)\big]\Big\}& . \end{eqnarray}
This result agrees precisely with the terms in (2.26) of Ref.\
\cite{berends}
through order $\alpha^2L$.

We illustrate the agreement we have found above in Figs. 2-5, for the
case \hbox{$f\bar f = \mu^- \mu^+$}.
\begin{figure}
\begin{center}
   \IfFileExists{beta1-01-fg1c.eps}{
	\epsfig{file=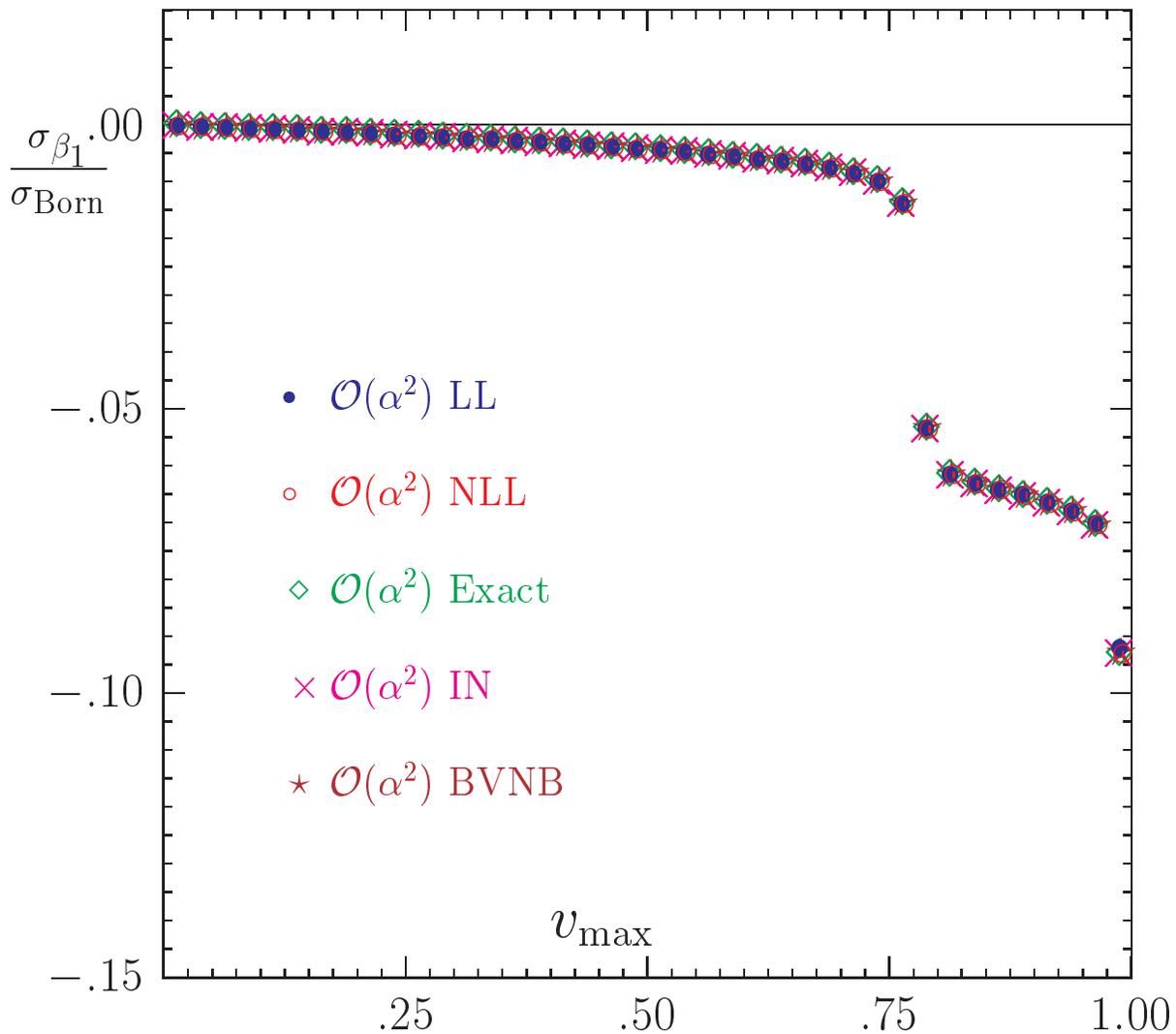,width=16cm}
	}{
	\IfFileExists{beta1-01-fg1.eps}{
  		\epsfig{file=beta1-01-fg1.eps,width=16}
		}{
		\message{Skipping <beta1-01-fg1c.eps>}
		\hbox{[ Figure beta1-01-fg1c.eps not included ]}
		}
	}
\end{center}
\vspace{-5mm}
\footnotesize\sf
\caption{
This is the $\beta_1^{(2)}$ distribution for the YFS3ff MC
(YFS3ff is the EEX3 matrix element option of the \KKMC\ in 
Ref.\ \protect\cite{kkmc:2001}), as a function of energy 
cut $v_{\max}$. It is divided by the Born cross-section.    
The IN result is from Ref.\ \protect\cite{in:1987}, and the BVNB result
is from Ref.\ \protect\cite{berends}.
}
\label{fig:Figs-B1}
\end{figure}
In Fig.\ \ref{fig:Figs-B1}, we show the complete $\beta_1^{(2)}$ distribution
for our exact result, our NLL and LL approximate results,
the result of Igarashi and Nakazawa {\it et al.}~\cite{in:1987}, 
and the result of Berends {\it et al.}~\cite{berends}. 
What we see is that there is a very good general agreement between all of 
these results.  To better assess the difference between them, we plot in 
Fig.\ \ref{fig:Figs-Bdif} the difference between the respective
${\cal O}(\alpha^2)$ and ${\cal O}(\alpha^1)$ results.
Again we see very good agreement except for the hardest possible
photons, where then the $LL$ result differs significantly from the others.
\begin{figure}
\begin{center}
   \IfFileExists{beta1-01-fg2c.eps}{
  	\epsfig{file=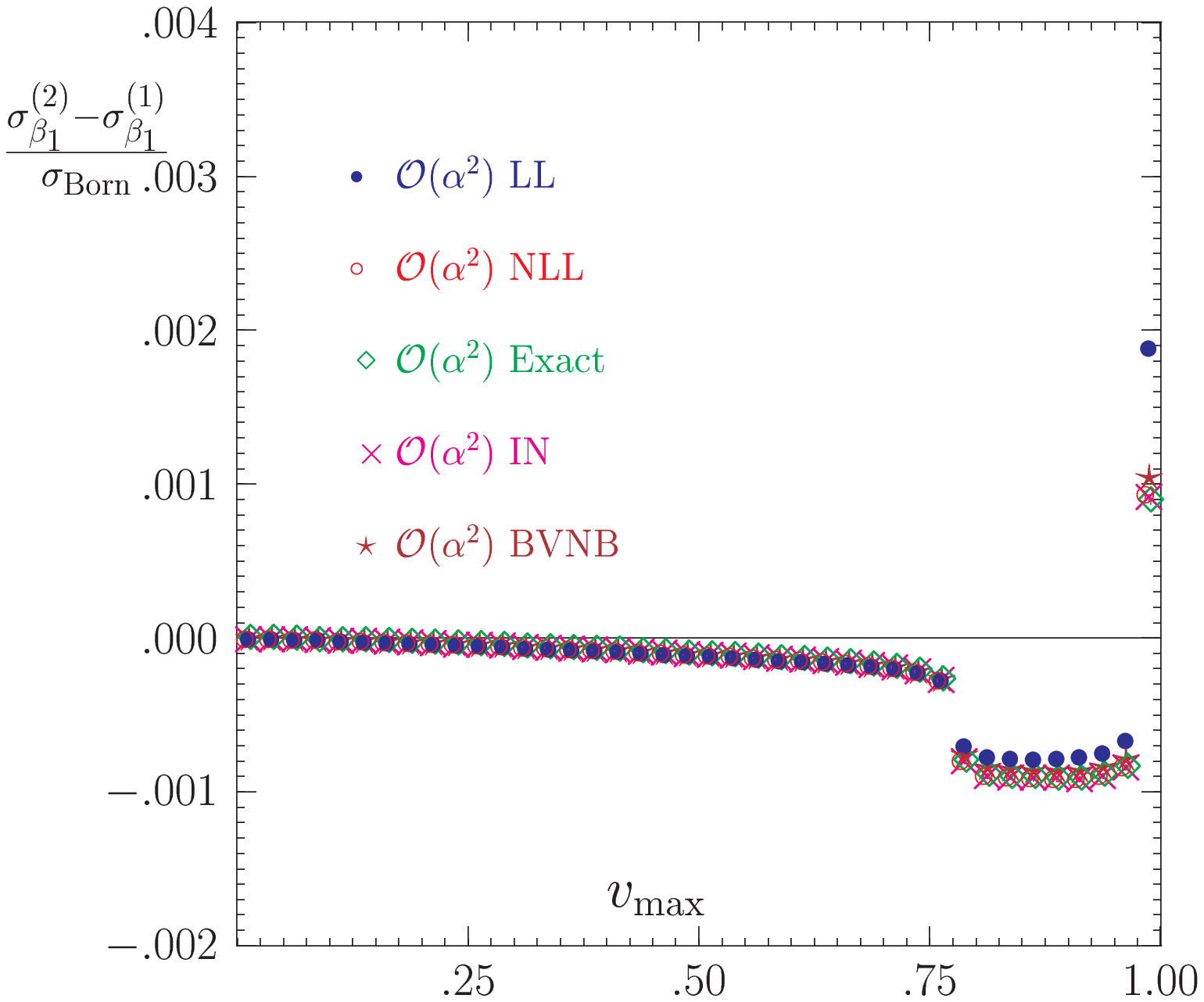,width=16cm}
	}{
	\IfFileExists{beta1-01-fg2.eps}{
  		\epsfig{file=beta1-01-fg2.eps,width=16cm}
		}{
		\message{Skipping <beta1-01-fg2c.eps>}
		\hbox{[ Figure beta1-01-fg2c.eps not included ]}
		}
	}
\end{center}
\vspace{-5mm}
\footnotesize\sf
\caption{
Difference $\beta_1^{(2)} - \beta_1^{(1)}$ for the YFS3ff MC
(the EEX3 option in the \KKMC), as a              
function of the cut $v_{\max}$. It is divided by the Born cross-section.       
}
\label{fig:Figs-Bdif}
\end{figure}

\begin{figure}
\begin{center}
   \IfFileExists{beta1-01-fg3c.eps}{
  	\epsfig{file=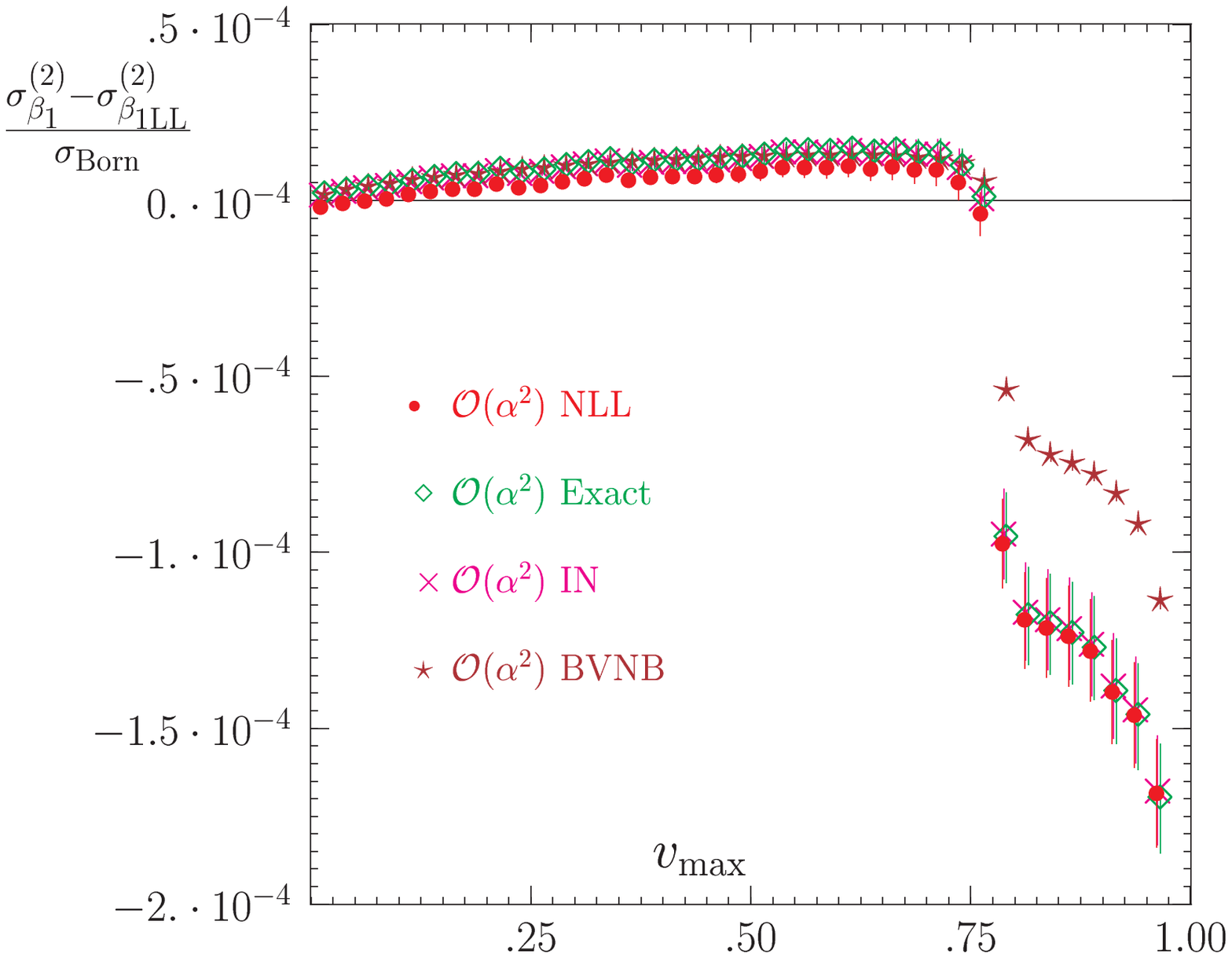,width=16cm}
	}{
	\IfFileExists{beta1-01-fg3.eps}{
  		\epsfig{file=beta1-01-fg3.eps,width=16cm}
		}{
		\message{Skipping <beta1-01-fg3c.eps>}
		\hbox{[ Figure beta1-01-fg3c.eps not included ]}
		}
	}
\end{center}
\vspace{-5mm}
\footnotesize\sf
\caption{
Next-to-leading-log contribution $\beta_1^{(2)} - \beta^{(2)}_{1 \rm LL}$ 
for the YFS3ff MC (the EEX3 option of the \KKMC), as a 
function of the cut $v_{\max}$. It is divided by the Born cross-section.       
}
\label{fig:Figs-BNLL}
\end{figure}

To isolate the respective predictions for the NLL effect,
we plot in Fig.\ \ref{fig:Figs-BNLL} the respective differences
between our LL ${\cal O}(\alpha^2)$ result and the 
other four results. We see that there is again very good agreement
but, at the level of $0.5\cdot10^{-4}$, the result of Ref.~\cite{berends}
is somewhat smaller in magnitude than the other three NLL results
in the $Z$ radiative return regime above a cut of 0.75. 

\begin{figure}
\begin{center}
   \IfFileExists{beta1-01-fg4c.eps}{
        \epsfig{file=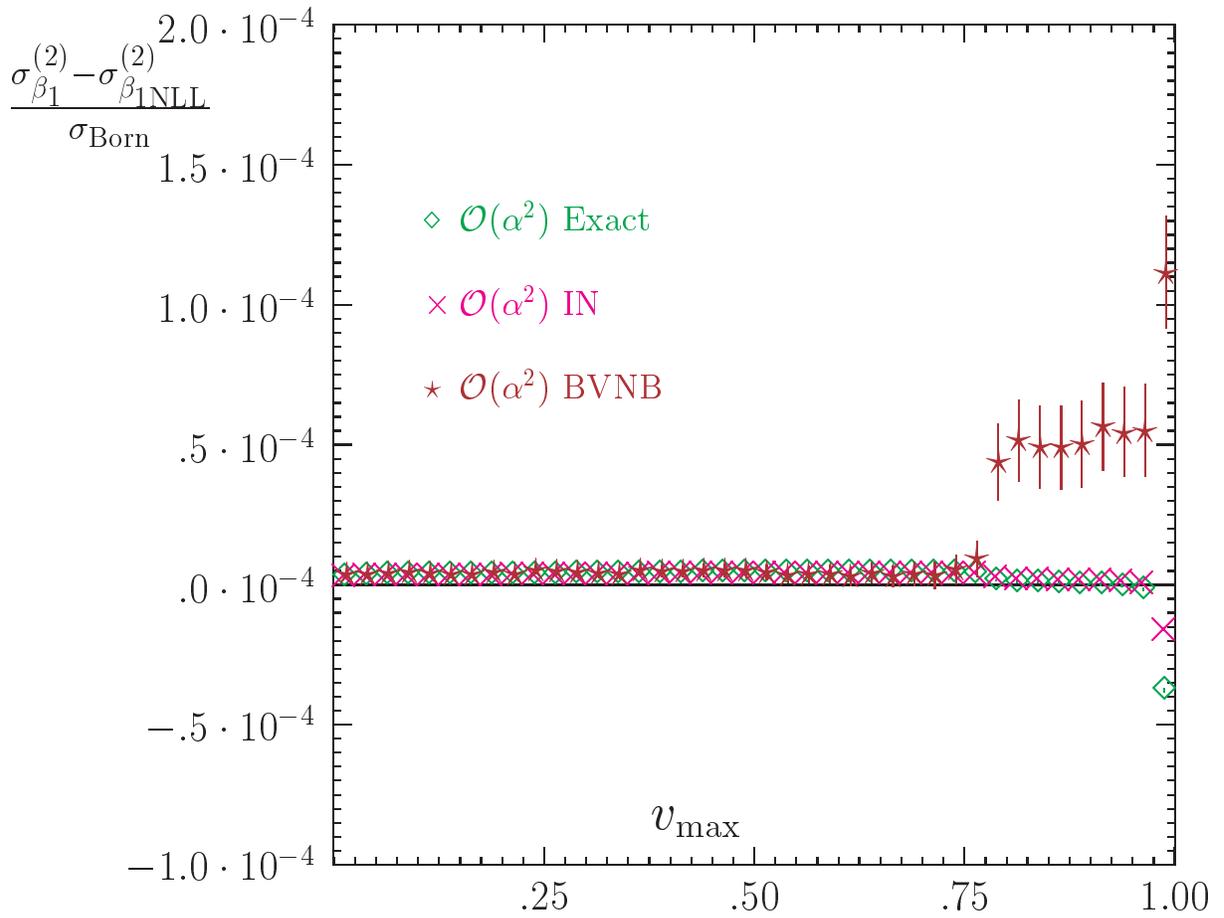,width=16cm}
	}{
	\IfFileExists{beta1-01-fg4.eps}{
  		\epsfig{file=beta1-01-fg4.eps,width=16cm}
		}{
		\message{Skipping <beta1-01-fg4c.eps>}
		\hbox{[ Figure beta1-01-fg4c.eps not included ]}
		}
	}
\end{center}
\vspace{-5mm}
\footnotesize\sf
\caption{
Sub-NLL contribution $\beta_1^{(2)} - \beta^{(2)}_{1 \rm NLL}$ for the 
YFS3ff MC (the EEX3 option of the \KKMC), as a function of the cut 
$v_{\max}$. It is divided by the Born cross-section.  
}
\label{fig:Figs-BNNLL}
\end{figure}

Finally, in the Fig.\ \ref{fig:Figs-BNNLL} we isolate the size of the
three NNLL results by plotting the difference between our NLL result
and our exact result, the result from Ref.~\cite{in:1987} and the
result from Ref.~\cite{berends}. We again see that the most pronounced
difference in the results occurs for the regime above $v_{max}=0.75$
where the result of Ref.~\cite{berends} differs by $0.5\cdot10^{-4}$
from the other two for $v_{max}<0.975$ and differs from the 
exact result by $1.5\cdot10^{-4}$ for $v_{max}>0.975$. The result
from Ref.~\cite{in:1987} differs from the exact result by 
$0.2\cdot10^{-4}$ for $v_{max}>0.975$
but is essentially indistinguishable from it for
smaller values of $v_{max}$.
We conclude that our exact result for the ${\cal O}(\alpha^2)$
correction $\bar\beta_1^{(2)}$ has a total precision tag of
$1.5\cdot10^{-4}$. Its NLL effect has already been implemented in the
\KKMC\ in Ref.~\cite{kkmc:2001}.

We have also made the analogous study to Figs. 2-5 for 500 GeV. 
We find very similar results, with the total precision tag of $2\cdot 10^{-4}$.

\section{Conclusions}
\label{conclusions}
\setcounter{equation}{0}

In this paper, we have presented exact results for the 
virtual correction to the process $e^+e^-\rightarrow f\bar f +\gamma$
for the $ISR\oplus FSR$. The results
are already in use in 
the \KKMC\ in Ref.~\cite{kkmc:2001} in connection
with the final LEP2 data analysis.

We have compared our results with those in Refs.~\cite{in:1987,berends}
and in general we find very good agreement, both at 200 GeV and at 500 GeV. 
For example, the size of
the NNLL correction is shown to be at or below the level of $2\cdot10^{-4}$
for all values of the energy cut parameter. Our results are
fully differential and are therefore ideally suited for MC 
event generator implementation. This has been done in the \KKMC\ in 
Ref.~\cite{kkmc:2001}. To compare our results with the results in
Refs.~\cite{berends}, we have partially integrated them accordingly.
While the results in Ref.~\cite{in:1987} are also fully differential,
they lack the complete mass corrections that our results do have.
In this way, one sees that our results are in fact unique.
They are an important part of the complete ${\cal O}(\alpha^2)$ corrections
to the $2f$ production process needed for precision
studies of such processes in the final LEP2 data analysis and in the
future TESLA/LC physics.

\section*{Acknowledgments}
\label{acknowledgments}

Two of the authors (S.J.\ and B.F.L.W.) would like to thank 
Prof.\ G.\ Altarelli of the CERN TH Div.\ and Prof. D. Schlatter
and the ALEPH, DELPHI, L3 and OPAL Collaborations, respectively, 
for their support and hospitality while this work was completed. 
B.F.L.W.\ would like to thank Prof.\ C.\ Prescott of Group A at SLAC for his 
kind hospitality while this work was in its developmental stages.

\section*{Note Added}
After we had submitted this paper, we became aware of related work by
G. Rodrigo, A. Gehrmann-De Ridder, M. Guilleaume and J. H. Kuhn,
hep-ph/0106132. These authors also agree with the analogous results
of Ref.~\cite{berends}, when the photon azimuthal angle is integrated
and the photon polarization is summed for the ISR process.

\appendix
\section{Scalar Integrals}
\renewcommand{\theequation}{A.\arabic{equation}}
\setcounter{equation}{0}

Previously, in Ref.~\cite{virtual}, the analogous exact virtually 
corrected photon cross sections were expressed in the $t$ channel using 
scalar integral functions which were calculated by a numerical package 
FF described in Ref.~\cite{scalar}. 
In the present paper, we have expressed these functions directly in terms of
logarithms and dilogarithms. This appendix will give the expressions for
the individual scalar integrals in the $s$ channel.

The notation for the scalar integrals will match Ref.~\cite{virtual}, with
kinematic notation defined in Sections \ref{preliminaries} and 
\ref{exact-virtual} of this paper.  The scalar integrals with two denominators 
are denoted $B$, and the ones appearing in the form factors are
\begin{eqnarray}
B_{12} &=& B(m_e^2; m_\gamma, m_e) , \nonumber\\
B_{13}^{r_i} &=& B(m_e^2-sr_i; m_\gamma, m_e) , \nonumber\\
B_{23} &=& B(m_\gamma^2; m_e, m_e) , \nonumber\\
B_{24} &=& B(s; m_e, m_e) , \nonumber\\
B_{34} &=& B(s'; m_e, m_e) ,
\end{eqnarray}
where the first argument is the square of the momentum through the diagram,
and the remaining arguments are the masses of the two lines. These functions
are UV divergent, but only the following finite combinations are needed:
\begin{eqnarray}
\label{B13dif}
B_{13}^{r_i} - B_{34} &=& \ln{s'\over sr_i} 
        + {m_e^2\over m_e^2-sr_i}\ln{sr_i\over m_e^2} - i\pi , \\
\label{B24dif}
B_{24} - B_{34} &=& \ln {s'\over s} , \\
\label{B12dif}
B_{12} - B_{34} &=& \ln {s'\over m_e^2} - i\pi .
\end{eqnarray}
The mass term in \eqn{B13dif} has been dropped when applying this expression,
since the mass corrections are added explicitly to the massless limit of the
calculation using the prescription of Ref.~\cite{berklei}. 

The integrals for three and four denominators are obtained from the appendix 
of Ref.~\cite{in:1987}. For three denominators, we need the expressions
\begin{eqnarray}
C_{123}^{r_i} &=& C(m_e^2, m_\gamma^2, m_e^2-sr_i; m_\gamma, m_e, m_e) ,
    \nonumber\\
C_{134}^{r_i} &=& C(m_e^2-sr_i, s', m_e^2; m_\gamma, m_e, m_e) , \nonumber\\
C_{234} &=& C(m_\gamma^2, s', s; m_e, m_e, m_e) , \nonumber\\
C_{124} &=& C(m_e^2, s, m_e^2; m_\gamma, m_e, m_e) , \nonumber
\end{eqnarray}
where the first three arguments are the squares of the external momenta,
and the next three are the masses of the three lines, in cyclic order.
The results are
\begin{eqnarray}
\label{C123}
sr_i C_{123}^{r_i} &=& -{1\over2}\ln^2{m_e^2\over sr_i} - 
        \Sp\left(1 - {m_e^2\over sr_i}\right) - {\pi^2\over 6} , \\
\label{C134}
(1 - r_j)s\,C_{134}^{r_i} &=& {1\over2} \ln^2{s'\over m_e^2}
    - {1\over2}\ln^2{(1-r_j)s\over m_e^2} - {1\over2}\ln^2{1-r_j\over r_i}
    \nonumber\\
    &+& 2\ln{s'\over m_e^2} \ln\left({1-r_j\over r_i}\right)
     + \ln{m_e^2\over sr_i} \ln\left({1-r_j\over r_i}\right)
     - \Sp\left({m_e^2\over sr_i}\right) \nonumber\\
    &-& 2\Sp\left({s'\over s(1-r_j)}\right) - {5\pi^2\over 6}
     - \pi i \ln{s'\over m_e^2} - 2\pi i \ln\left({1-r_j\over r_i}\right) ,\\
\label{C234}
(r_1 + r_2)s\,C_{234} &=& {1\over 2} \ln^2 {s\over m_e^2}
        - {1\over 2} \ln^2{s'\over m_e^2} + \pi i \ln{s'\over s} , \\
\label{C124}
s C_{124} &=& {1\over 2}\ln^2{s\over m_e^2} - \ln{s\over m_e^2}
        \ln{m_\gamma^2\over m_e^2} - {2\pi^2\over 3} 
        + \pi i \ln{m_\gamma^2\over s}.
\end{eqnarray}
The expression \eqn{C123} is an analytic continuation of the one in 
Ref.~\cite{in:1987}, as required since $r_i < m_e^2$ is possible, 
and the expression \eqn{C134} drops mass terms. 
In particular, the photon mass regulator $m_\gamma$ is dropped whenever
possible. 

The expression with four denominators is 
\begin{eqnarray}
\label{D1234}
s^2 r_i D_{1234}^{r_i} &=& \ln^2{s'\over m_e^2} 
	- 2\ln{s\over m_e^2}\ln{s r_i\over m_e^2}
    + \ln{m_\gamma^2\over m_e^2}\left(\ln{s\over m_e^2} - i\pi\right)
    \nonumber\\
    &+& 2\,\Sp(r_1+r_2) - {5\pi^2\over 6} - 2\pi i\ln{s'\over sr_i} .
\end{eqnarray}
All of these expressions have been checked for agreement with the FF package.
The function $R$ defined in equations \eqn{Ra2} and \eqn{Rb2} and appearing
in the virtual photon factors is the IR-finite combination
\begin{equation}
R(r_i, r_j) = s\,(C_{124} + sr_j D_{1234}^{r_j}) - sr_j C_{123}^{r_j}
        - (1-r_i)s\,C_{134}^{r_j} + (r_1 + r_2)s\,C_{234} . 
\end{equation}
This completes the appendix.

\bibliographystyle{unsrt}

\section*{Figure Captions}
\label{figure-captions}

\par\noindent
Fig.\ \ref{fig:graphs}. 
Initial state radiation graphs for $e^+ e^- \rightarrow f\overline{f}$
with one virtual and one real photon, for $f\neq e$.

\par\noindent\\[2pt]
Fig.\ \ref{fig:Figs-B1}.
This is the $\beta_1^{(2)}$ distribution for the YFS3ff MC
(YFS3ff is the EEX3 matrix element option of the \KKMC\ in 
Ref.\ \cite{kkmc:2001}), as a function of energy cut $v_{\max}$. 
It is divided by the Born cross-section.

\par\noindent\\[2pt]
Fig.\ \ref{fig:Figs-Bdif}. 
Difference $\beta_1^{(2)} - \beta_1^{(1)}$ for the YFS3ff MC
(the EEX3 option in the \KKMC), as a function of the cut $v_{\max}$. 
It is divided by the Born cross-section.

\par\noindent\\[2pt]
Fig.\ \ref{fig:Figs-BNLL}. 
Next-to-leading-log contribution $\beta_1^{(2)} - \beta^{(2)}_{1 \rm LL}$ 
for the YFS3ff MC (the EEX3 option of the \KKMC), as a function of the cut 
$v_{\max}$. It is divided by the Born cross-section.

\par\noindent\\[2pt]
Fig.\ \ref{fig:Figs-BNNLL}. 
Sub-NLL contribution $\beta_1^{(2)} - \beta^{(2)}_{1 \rm NLL}$ for the          YFS3ff MC (the EEX3 option of the \KKMC), as a function of the cut 
$v_{\max}$. It is divided by the Born cross-section.

\end{document}